\documentclass[preprint]{elsarticle}
\usepackage{graphicx}
\usepackage{hyperref}

\begin{document}

\title{The MFA ground states for the extended Bose-Hubbard model with a three-body constraint}

\author[urfu]{Yu.D.~Panov\corref{cor1}}
\ead{yuri.panov@urfu.ru}
\author[urfu]{A.S.~Moskvin}
\author[urfu]{E.V.~Vasinovich}
\author[urfu]{V.V.~Konev}

\cortext[cor1]{Corresponding author}

\address[urfu]{Ural Federal University, 19 Mira str., 620002, Ekaterinburg, Russia}

\begin{abstract}
We address the intensively studied extended bosonic Hubbard model (EBHM) with truncation of the on-site Hilbert space to the three lowest occupation states $n=0,1,2$ in frames of the $S=1$ pseudospin formalism. 
Similar model was recently proposed to describe the charge degree of freedom in a model high-T$_c$ cuprate with the on-site Hilbert space reduced to the three effective valence centers, nominally Cu$^{1+;2+;3+}$. 
With small corrections the model becomes equivalent to a strongly anisotropic $S=1$ quantum magnet in an external
magnetic field. 
We have applied a generalized mean-field approach and quantum Monte-Carlo technique for the model 2D $S=1$ system with a two-particle transport to find the ground state phase with its evolution under deviation from half-filling.
\end{abstract}



\maketitle

\section{Introduction}

These days spin algebra and spin Hamiltonians are used not only in the traditional fields of spin magnetism  but in so-called pseudospin lattice systems with the on-site occupation constraint. 
For instance, the $S=1$ pseudospin formalism was applied to study an extended Bose-Hubbard model (EHBM) with truncation of the on-site Hilbert space to the three lowest occupation states $n=0,1,2$ (semi-hard-core bosons) considered to be three pseudospin states with $M={-}1$, $M=0$, $M={+}1$, respectively (see \cite{Moskvin15} and references therein). 
At variance with quantum $s=1/2$ systems the Hamiltonian of $S=1$ spin lattices in general is characterized by several additional terms such as a single ion anisotropy that results in their  rich phase diagrams.
Recently we made use of the $S=1$ pseudospin formalism to describe the charge degree of freedom in high-T$_c$ cuprates with the on-site Hilbert space reduced to only the three effective valence centers [CuO$_4$]$^{7-,6-,5-}$ (nominally Cu$^{1+;2+;3+}$)\,\cite{Moskvin11,Moskvin13,Moskvin15-SCES,Moskvin17}.

\section{$S=1$ (pseudo)spin Hamiltonian}
The $S=1$ spin algebra includes the eight nontrivial independent spin operators:  spin-dipole moment ${\bf S}$ and five spin-quadrupole operators $Q_{ij}=(\frac{1}{2}\{S_{i},S_{j}\}-\frac{2}{3}\delta_{ij})$ whose mean values define so-called spin-nematic order. 
Spin operators $S_{\pm}$ and $T_{\pm}=\{S_z, S_{\pm}\}$  change the pseudospin projection (and occupation number) by $\pm 1$, while $S_{\pm}^2$  changes the pseudospin projection by $\pm 2$.

Hereafter in the paper we will focus on a simplified 2D $S=1$ (pseudo)spin Hamiltonian with the nearest neighbor coupling and the  only  two-particle transport term (inter-site biquadratic anisotropy) as follows:
\begin{equation}
		\hat{H} 
		=  \sum_{i}  (\Delta S_{iz}^2
		- \mu S_{iz}) 
		+ V\sum_{\langle ij \rangle} S_{iz}S_{jz}
		- t\sum_{\langle ij \rangle}(S_{i+}^{2}S_{j-}^{2}+S_{i-}^{2}S_{j+}^{2})
		,
\label{Hsimple}	
\end{equation}
where $V>0$, $t>0$.
The first single-site term in $\hat{H}$ describes the effects of a bare pseudo-spin splitting and relates with the on-site density-density interactions, or correlations: $\Delta=U/2$. 
The second term, or a pseudospin Zeeman coupling may be related with a   pseudo-magnetic field $\parallel$\,$Z$  which acts as analog of chemical potential $\mu$ for doped charge with a charge constraint:
\begin{equation}
\sum _{i} \langle S_{iz} \rangle =  nN
,
\label{cc}
\end{equation}
where fixed $n$ is the doped charge density.
The third (Ising) term in $\hat{H}$ describes the effects of the short-  and long-range inter-site density-density interactions. The last term in ${\hat H}$ describes the two-particle inter-site  hopping.
In the strong on-site attraction limit of the model (large easy-axis pseudospin on-site anisotropy) we arrive at the Hamiltonian of the hard-core, or local, bosons which was earlier considered to be a starting point for explanation of the cuprate high-T$_c$ superconductivity\,\cite{Micnas1990}.
The spin counterpart of ${\hat H}$ corresponds to an anisotropic $S=1$ magnet with a single ion (on-site) and two-ion (bilinear and biquadratic) symmetric anisotropy in an external magnetic field. It describes an interplay of the Zeeman,   single-ion and two-ion anisotropic terms giving rise to a competition of an (anti)ferromagnetic  order along $Z$-axis with an in-plane $XY$ spin-nematic order.
A remarkable feature of the Hamiltonian (\ref{Hsimple}) is that the on-site pseudospin states $M=0$ and $|M|=1$ do not mix under the inter-site coupling. 
The model allows us to directly study a continuous transformation of the semi-hard-core bosons to the effective hard-core bosons formed by boson pairs  under driving the correlation parameter $\Delta=U/2$ to large negative values  ("negative-$U$ model"). 
The simplified model can be directly applied to a description of bosonic systems with suppressed one-particle hopping.

\section{Mean-field approximation}

To analyze the simplified model we start with a mean-field approximation (MFA) for 2D square lattice, however, at variance with a conventional classical MFA we made use of more correct approach that takes into account the quantum nature of the $S=1$ (pseudo)spin states\,\cite{Nadya}.
First we introduce a set of the on-site $S=1$ coherent states
\begin{equation}
|{\bf c}\rangle = c_{-1}|-1\rangle +c_0|0\rangle +c_{+1}|+1\rangle 
,
\label{function}
\end{equation}
where the $c_M$ coefficients can be represented as follows
\begin{equation}
	c_{1} {=} \sin\frac{\theta}{2}\cos\frac{\phi}{2} e^{-i\frac{\alpha}{2}}
	,\quad
	c_{0} {=} \cos\frac{\theta}{2} e^{i\frac{\beta}{2}}
	,\quad
	c_{-1} {=} \sin\frac{\theta}{2}\sin\frac{\phi}{2}e^{i\frac{\alpha}{2}}
\label{CoefWFinCB}
\end{equation}
with $\theta$, $\phi$, $\alpha$, $\beta$ to be parameters defined by the minimization of the energy.
The MFA energy can be written as follows
\begin{eqnarray}
	E 
	&=& 
	\frac{\Delta}{2} \sum_i \Big( 1 - \cos\theta_i \Big) 
	{} + \frac{V}{4} \sum_{\left\langle ij\right\rangle}  
	\Big( 1 - \cos\theta_i \Big) \Big( 1 - \cos\theta_j \Big) \cos\phi_i \, \cos\phi_j
	-{} 
	\nonumber
	\\[0.5em]
	&&
	{}	- \frac{t}{8} \sum_{\left\langle ij\right\rangle} 
	\Big( 1 - \cos\theta_i \Big) \Big( 1 - \cos\theta_j \Big) \sin\phi_i \, \sin\phi_j \, \cos(\alpha_i-\alpha_j)
	-{} 
	\label{EnThetaPhi}
	\\[0.5em]
	&&
	{} - \frac{\mu}{2} \sum_i \Big( 1 - \cos\theta_i \Big) \cos\phi_i
	.
	\nonumber
\end{eqnarray}
Here, the term with the chemical potential $\mu$ takes into account the constraint (\ref{cc}).
It is worth noting that due to the absence of the one-particle inter-site hopping terms in Hamiltonian (\ref{Hsimple}) the energy does not depend on phase parameter $\beta$, so the $\beta$ remains undetermined.


\begin{table}
	\centering
\begin{tabular}{llll}
	\hline
	\phantom{\bigg|} 
	& $ \varepsilon $
	& $ \cos\theta_j $ 
	& $ \cos\phi_j $ 
	\\[0.25em] \hline
	&&&
	\\[-0.75em]
	SF 
	& $ \delta - 1 + n^2 (2\nu + 1) $
	& $ -1 $ 
	& $ n $
	\\[0.5em]
	SS 
	& $ \delta {-} 2\nu {+} 2 |n| \sqrt{4\nu^2 {-} 1} $
	& $ -1 $ 
	& $ n {+} (-1)^j a b $ 
	\\[0.5em]
	CO1 
	& $ |n| \, \delta $
	& $ 1  -  2 |n|  +  2 (-1)^j n $ 
	& $ \sigma $ 
	\\[0.5em]
	CO2 
	& $ \left(1-|n|\right) \delta + 4\left(|n|-\frac{1}{2}\right) \nu $
	& $ -1  +  2 |n|  +  2 (-1)^j n $ 
	& $ (-1)^{j+1} $ 
	\\[0.5em]
	CO3 
	& $ |n| \delta + 4\left(|n|-\frac{1}{2}\right) \nu $
	& $ 1  -  2 |n|  +  2 (-1)^j \sigma \big( 1 - |n| \big) $ 
	& $ \sigma $
	\\[0.5em]  \hline
\end{tabular}
\caption{
The energies and parameters of MFA GS phases. The index $j$ is 0(1) for A(B) sublattice. The details of notations see in the text.
}
\label{table:En}
\end{table}


\begin{table}
	\centering
\begin{tabular}{llll}
	\hline
	\phantom{\bigg|} 
	& $ \langle S_{z} \rangle_j $ 
	& $ \langle P_0 \rangle_j $ 
	& $ \langle S_{\pm}^2 \rangle_j $ 
	\\[0.25em] \hline
	&&&
	\\[-0.75em]
	SF 
	& $ n $ 
	& 0 
	& $ \frac{\eta}{2} \sqrt{1-n^2} \, e^{\pm i\alpha} $ \\[0.5em]
	SS 
	& $ n + (-1)^j a b $ 
	& 0 
	& $ \frac{\eta}{2} \big( a{-} (-1)^j \sigma \, b \big) \sqrt{|n|} \, e^{\pm i\alpha} $ \\[0.5em]
	CO1 
	& $ n - (-1)^j |n| $ 
	& $ 1 - |n| + (-1)^j n $ 
	& 0 \\[0.5em]
	CO2 
	& $ n - (-1)^j \big( 1 {-} |n| \big) $ 
	& $ |n| + (-1)^j n $ 
	& 0 \\[0.5em]
	CO3 
	& $ n - (-1)^j \big( 1 {-} |n| \big) $ 
	& $ \big( 1 {-} |n| \big) \big( 1 {+} (-1)^j \sigma \big) $
	& 0 \\[0.5em]  \hline
\end{tabular}
\caption{
The order parameters of GS MFA phases. 
The index $j$ is 0(1) for A(B) sublattice. 
Here $\eta=\pm1$ and the phase $\alpha$ remains undefined.
Other details of notations see in the text.
}
\label{table:OrderPar}
\end{table}


In a two-sublattice A-B model, we arrive at the five MFA uniform phases for the ground state (GS). 
The energies and parameters of solutions are listed in Table.\ref{table:En}. 
We use the notations:
$\varepsilon = E / (t N)$, 
$\delta = \Delta /t$,  
$\nu = V /t$, 
$\sigma = \mathop{\rm sgn} n$, 
$a = \sqrt{  \sqrt{ \frac{2\nu+1}{2\nu-1} } - |n| } $, 
$b = \sqrt{  \sqrt{ \frac{2\nu-1}{2\nu+1} } - |n| } $. 
In all phases, the value of 
chemical potential $\mu$ satisfies the regular expression $\mu = t \, \partial \varepsilon / \partial n$.
The solutions for SF and SS phases imply that $\alpha_A - \alpha_B = 0 \mbox{ or } \pi$, 
in other phases this difference remains undefined.

The GS MFA phases differ by local charge density $\langle S_{z} \rangle$ 
and local density of $M=0$ (Cu$^{2+}$) states $\langle P_{0} \rangle = 1 - \langle S_{z}^2 \rangle$:
\begin{equation}
	\left\langle S_{z} \right\rangle
	=
	\frac{1}{2} \left( 1 - \cos\theta \right) \cos\phi
	,\quad
	\left\langle P_{0} \right\rangle
	=
	\frac{1}{2} \left( 1 + \cos\theta \right)
	,
\end{equation}
and by local superfluid order parameter, or pseudospin nematic order $\langle S_{\pm}^2 \rangle$:
\begin{equation}
	\left\langle S_{\pm}^2 \right\rangle
	=
	\frac{1}{4} \left( 1 - \cos\theta \right) \sin\phi \; e^{\pm i \alpha}
	.
\end{equation}
The density of superfluid component is related to helicity modulus \cite{Fisher1973}. 
This allow us to find an expression of the superfluid density $\rho$ in terms of local superfluid order parameters in the two-sublattice MFA:
\begin{equation}
	\rho = \mathop{\mathrm{Re}} \big( \left\langle S_{A+}^2 \right\rangle \! \left\langle S_{B-}^2 \right\rangle \big)
	.
\end{equation}
The local order parameters for the GS MFA phases are listed in Table.\ref{table:OrderPar}.

Bose superfluid (SF) and supersolid (SS) phases are completely analogous to phases of charged hard-core bosons 
\cite{Micnas1990,Schmid2002} as these phases have no the $M = 0$ states. 
The superfluid density in SF phase, $\rho = (1-n^2)/4$, 
has maximum value at $n=0$ and does not depend on inter-site density-density interactions parameter $\nu$. 
In SS phase, the superfluid density  $\rho = |n|/(2\sqrt{4\nu^2 - 1})$ decreases with rising of $\nu$. 
The charge density differs on sublattices in SS phase  and this phase becomes the pure charge-ordered one at $n=0$.

Stability conditions for SF phase 
\begin{equation}
	\delta < 2
	,\quad
	n^2 > \frac{2\nu-1}{2\nu+1}
	,
	\label{stabcondSF}
\end{equation}
and for SS phase
\begin{equation}
	\delta < 2
	,\quad
	\nu > \frac{1}{2}
	,\quad
	\sqrt{\frac{2\nu-1}{2\nu+1}} - \frac{\left(1-\frac{\delta}{2}\right)^2}{\delta\sqrt{4\nu^2 - 1}}
	< |n| 
	< \sqrt{\frac{2\nu-1}{2\nu+1}}
	,
	\label{stabcondSS}
\end{equation}
define the boundary expression for SF and SS phases: $n^2 = (2\nu-1)/(2\nu+1).$
As the energies of SF and SS phases have the same dependence on the correlation parameter $\delta$ (see Table \ref{table:En}), 
the line of the SF-SS transition does not change with $\delta$.

Three charge ordered MFA phases
with $\left\langle S_{A,B\pm}^2 \right\rangle = 0$ but different types of the sublattice occupation
emerge if $\delta>0$ and completely displace the superfluid phases at $\delta>2$.

Stability conditions for the charge ordered 1 (CO1) are given by inequality
\begin{equation}
	|n| < \min \left\{  \frac{1}{2} , \, \frac{\delta}{4\nu}  \right\}
	.
	\label{stabcondCO1}
\end{equation}
Given $n = 0$ the CO1 phase consists of $M=0$ centers. 
The striking feature of the CO1 phase is the independence of energy on inter-site interaction parameter $\nu$. 
According to the two sublattices mean field approach, 
upon doping only one of the sublattices begins to be filled by $M=\pm1$ centers depending on the sign of $n$. 
Numerical simulations with classical Monte-Carlo show that there is no difference in the sublattices occupations while $|n|\ll1/2$, 
but this difference arises at $|n|\rightarrow1/2$ 
according to the MFA expressions for $\langle S_{z} \rangle_j$ and $\langle P_0 \rangle_j$.

Charge ordered 2 (CO2) phase has the stability conditions given by the expression
\begin{equation}
	|n| < \min 
	\left\{ 
		\frac{1}{2} 	,\, 
		1-\frac{\delta}{4\nu} 	,\, 
		\frac{ 8 \delta \nu - \delta^2 - 4 }{ 8 | 1 - \delta \nu | } 
	\right\}
	.
	\label{stabcondCO2}
\end{equation}
At $n=0$, the CO2 phase is fully polarized, and with a deviation from $n=0$ one of the sublattices is filled by $M=0$ centers that leads to reducing of its $|\langle S_{z} \rangle|$.

The line of the CO1-CO2 transition for all $|n|<1/2$ is defined by the expression $\delta=2\nu$ that follows from the equality of energies of these phases.

Given $|n|=1/2$ the parameters of CO1 and CO2 phases become equal to that of the charge ordered 3 (CO3) phase. 
Stability conditions of the CO3 phase are given by
\begin{equation}
	\frac{1}{2} < |n| < \min \left\{ 1 ,\, \frac{ 8 \delta \nu + \delta^2 + 4 }{ 8 ( 1 + \delta \nu ) } \right\}
	.
	\label{stabcondCO3}
\end{equation}
For the CO3 phase at $n =1/2$, one of the sublattices is completely filled with $M =1 $ or $M = -1$ centers depending on the sign of $n$, while the second is completely filled by $M = 0$ centers. 
With the $|n|$ rising, the second sublattice is also filled by $|M|=1$ centers.

\begin{figure}
	\centering
		\includegraphics[width=\linewidth]{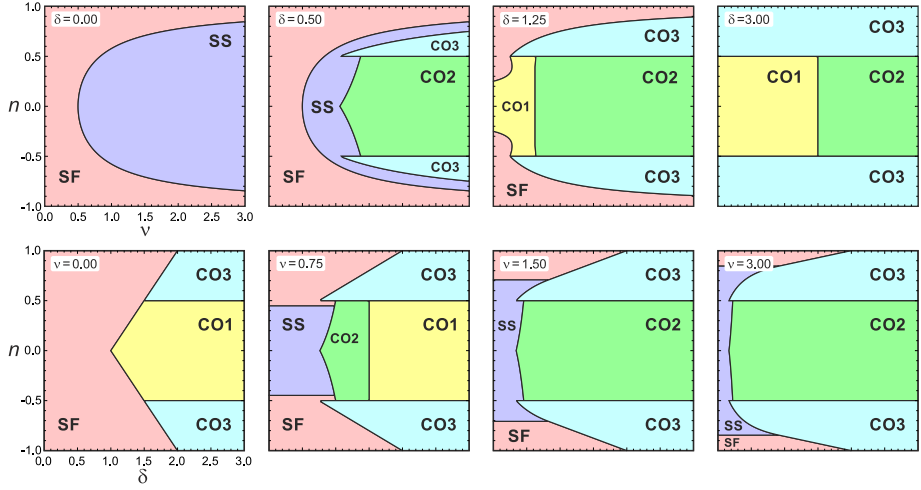}
	\caption{
	(Color online)
	The MFA GS phase diagrams for the inter-site interaction parameter $\nu$ variation (upper panels) and for the on-site correlation parameter $\delta$ variation (lower panels).
	}
	\label{fig:PD}
\end{figure}

Interestingly, all the local order parameters do not depend on the correlation parameter $\Delta$, while this parameter governs the energy of different phases. 
Taking into account the on-site correlations and the stability conditions (\ref{stabcondSF}--\ref{stabcondCO3}) we arrive at very rich and intricate phase diagrams for the model system as compared with relatively simple phase diagrams for hard-core bosons\,\cite{Micnas1990,Schmid2002}.
The kind of transition between the GS phases is determined by the limiting values of the order parameters (see Table \ref{table:OrderPar}) on the transition lines. 
The SF-SS transition does not lead to discontinuities of the order parameters (the transition of the second kind) except the jump of the local superfluid order parameter $\langle S_{\pm}^2 \rangle$ at $n=0$ (the point of the first kind transition). 
The CO1-CO3 and CO2-CO3 transitions at $n=1/2$ are also continuous (of the second kind). 
All other transitions are discontinuous (of the first kind).

In Fig.\,\ref{fig:PD} (upper panels) we show the MFA GS phase diagrams for the inter-site interaction parameter $\nu$ variation and for the on-site correlation parameter $\delta$ variation (lower panels).
For $\delta=0$, the phase diagram is the same as for hc-bosons \cite{Micnas1990}. 
With increasing $\delta$, superfluid phases are rapidly replaced with the charge ordered phases. 
The replacement of the SS phase begins at $\delta>0$ in the region of large values of the parameter $\nu$. 
SS phase disappears completely when $\delta\approx1.15$. 
For $\delta>1 $, in the region of small values of the parameter $\nu$, the CO1 phase appears, which begins to displace the SF phase.
This process begins at $n=0$, where the value of the density of the superfluid component is maximal.
For $\delta\geq2$, the SF phase is completely replaced with the charge ordered phases.
 
Evolution with a change in the parameter $\nu$ also shows a rapid decrease in the fraction of superfluid phases on the phase diagram in comparison with the charge ordered phases.
The most complicated phase diagram is  observed for $\delta\approx1.1$, $\nu\approx0.65$ where the competition of the on-site and  intersite interactions manifests itself most strongly.
At half-filling $n\,=\,0$ the positive values of the correlation parameter $\delta$ stabilize a limiting  CO1 phase 
with $\left\langle S_{A,Bz} \right\rangle = \left\langle S_{A,Bz}^2 \right\rangle\,=\,0$,  
or a ''parent Cu${}^{2+}$ phase'' for a model cuprate, 
while positive values of $\nu$ stabilize a limiting  CO2 phase 
with $\left\langle S_{A,Bz} \right\rangle =\pm 1$; $\left\langle S_{A,Bz}^2 \right\rangle\,=\,1$, 
or a checkerboard ''antiferromagnetic'' order of pseudospins along $z$-axis, 
or a disproportionated Cu${}^{1+}$-Cu${}^{3+}$ phase for a model cuprate. 
As a result of the competition between the on-site and inter-site correlations we arrive at a ''starting'' CO1 phase for $\delta>2\,\nu$ or CO2 phase for $\delta \leq 2\,\nu$. 
At $n\,=\,0.5$ we see a transformation of the CO1 and CO2 phases into the CO3 phase. 
The line of the first order phase transition CO3-SF in Fig.\,\ref{fig1} corresponds to the equality of the respective energies.  
It is worth to note that the critical concentration $n$ for the SS-SF, CO1-CO3 and CO2-CO3 transitions does not depend on the correlation parameter $\delta$.

In Fig.\,\ref{fig2} (top panel, solid lines) we present the $n$-dependence of the correlation functions 
$S_{zz}(\pi ,\pi)\,=\,\langle S_z,S_z\rangle$ (static structure factor) 
and $S^2_{+-}(0,0)\,=\,\langle S^2_+,S^2_-\rangle$ at $\delta\,=\,1.5$, $\nu\,=\,0.75$, 
determining the long-range  CO and SF orders, respectively, given $\delta=1.5$, 
that is in an immediate closeness to CO2-CO1 phase transition for small $n$.

\section{Quantum Monte-Carlo calculations}

\begin{figure}[t]
\centering
\includegraphics[width=0.75\linewidth]{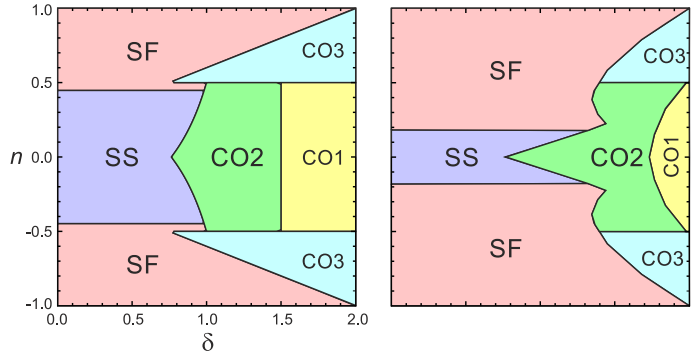}
\caption{
(Color online)
The  $n$\,-\,$\delta$ GS phase diagrams for the model system given $\nu=0.75$. Left panel shows the MFA results, right panel shows the QMC results. 
}
\label{fig1}
\end{figure}

\begin{figure}
\centering
\includegraphics[width=0.9\linewidth]{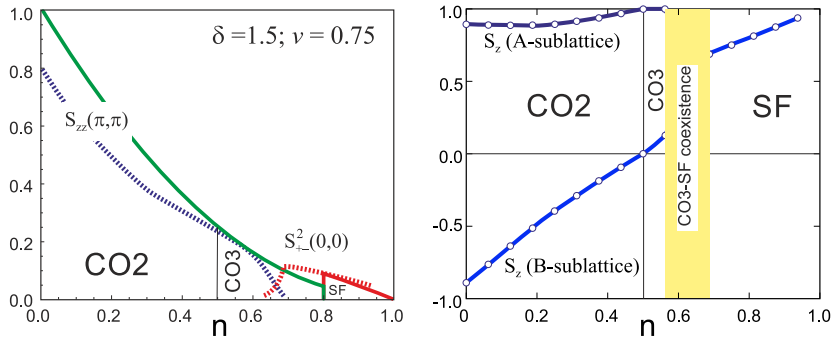}
\caption{
(Color online) 
Left panel: 
Correlation functions for the model $S=1$ pseudospin system given $\delta=1.5$,  $\nu=0.75$, 
solid lines are the MFA results, dotted lines are the QMC results. 
Right panel: 
QMC data for the sublattice $S_z$-components as functions of the deviation from the half-filling. 
Filling points to a CO3-SF coexistence phase typical for the first kind phase transition. 
}
\label{fig2}
\end{figure}

We have performed Quantum Monte-Carlo (QMC)\,\cite{Bauer} calculations for our model Hamiltonian (\ref{Hsimple}).
In Fig.\,\ref{fig1} we  compare the ground state $\delta$\,-\,$n$ phase diagram of our model 2D system calculated on square lattice 12\,$\times$\,12 given $\nu\,=\,0.75$ with that of calculated within MFA approach. 
As for a simple hard-core counterpart\,\cite{Micnas1990,Schmid2002}, despite some qualitative agreement, we see rather large quantitative difference between two diagrams in Fig.\,\ref{fig1}. 
In particular, it concerns  a clearly larger volume of the quantum SF phase that might be related with a sizeable suppression of quantum fluctuations within MFA approach. 
The SF-SS transition line does not depend on the correlation parameter $\delta$ in MFA calculations as well as in QMC ones since both these phases consist of only the $M=\pm1$ states having the same dependence of the energy on $\delta$. 
The location of the CO1-CO3 and CO2-CO3 transition lines at $|n| = 0.5$, both for MFA and QMC, has a trivial structural reason. 
The filling of the lattice by $M=\pm1$ centers for CO1 phase or by $M=0$ centers for CO2 phase during the doping leads on the lines $n=\pm0.5$ to identical result that minimizes the energy of the inter-site density-density interactions. 
Namely, this is the initial state of CO3 phase, when the first sublattice is completely filled by $M=0$ centers and the second one is completely filled by $M = \pm1$ centers. 
In contrast to MFA, the CO1-CO2 transition line in QMC calculations shows evident dependence on $n$ that implies a more complicated structure of the CO1 and CO2 phases as compared with MFA. 
This leads, in particular, to the fact that the triple point of the CO1-CO2-CO3 phases shifts 
from the MFA values $n=0.5$, $\delta=1.5$ to $n=0.5$, $\delta=2.0$.

In Fig.\,\ref{fig2} (left panel, two dotted lines) we present the QMC calculated  static structure factor $S_{zz}(\pi ,\pi)$ and the superfluid (pseudospin nematic) correlation function $S^2_{+-}(0,0)$. 
It is worth to note a semiquantitative agreement with the MFA data. 
Smaller value of the quantum structure factor $S_{zz}(\pi ,\pi)$ at $n\,=\,0$ is believed to be a result of the pseudospin reduction due to quantum fluctuations.
Right panel in Fig.\,\ref{fig2} shows the $n$-dependence of the mean sublattice $S_z$ values, $S_{Az}$ and $S_{Bz}$,
that clearly demonstrates the pseudospin quantum reduction effect within CO2 phase  and specific  features of the sublattice occupation, or ''pseudo-magnetization'' under CO2-CO3-SF transformation. 
Also, note that these QMC data points to the CO3-SF phase coexistence typical for the first kind phase transition, but obviously absent in MFA.

It should be noted that the results of QMC calculations for the system 12\,$\times$\,12 presented here vary slightly compared to the system 8\,$\times$\,8, that supports their validity. Calculations for larger lattices are in progress.

\section{Conclusions}

A simplified 2D $S=1$ pseudospin Hamiltonian with a two-particle transport term (pseudospin nematic coupling) was analyzed within a generalized MFA and QMC  technique. 
We have obtained the ground-state phase diagrams and correlation functions given different values of the coupling parameters with a focus on the role of the on-site correlation effect (single-ion anisotropy). 
The comparison of the two methods allows us to uncover fundamental shortcomings of the MFA technique and  clearly demonstrate the role of quantum effects.

\section{Acknowledgement}

The research was supported  by the Government of the Russian Federation, Program 02.A03.21.0006 and by the Ministry of Education and Science of the Russian Federation, projects Nos. 2277 and 5719.

\end{document}